\documentclass[10pt]{IEEEtran}
\usepackage[utf8]{inputenc}
\usepackage{epsfig}
\usepackage{graphicx}
\pdfoutput=1
\usepackage{color}
\usepackage{amsmath}
\usepackage{amssymb}
\usepackage{multirow}
\usepackage{array}
\usepackage{diagbox}
\usepackage{url}
\usepackage{makecell}
\usepackage{bbding}
\usepackage{pifont}
\usepackage[ruled,vlined,linesnumbered]{algorithm2e}

\title{A Scale-Arbitrary Image Super-Resolution Network Using Frequency-domain Information}

\author{Jing Fang\IEEEauthorrefmark{1}, Yinbo Yu\IEEEauthorrefmark{1}, \IEEEmembership{Member, IEEE,} Zhongyuan Wang, \IEEEmembership{Member, IEEE,} Xin Ding,  \\
Ruimin Hu, \IEEEmembership{Senior Member, IEEE}
\thanks{The research was supported in part by the National Natural Science Foundation of China (NSFC) under Grant 91738302, Grant 62202387, Grant 62071339. (\textit{Corresponding author: Ruimin Hu.)}}
\thanks{* Equal Contribution.}
\thanks{J. Fang, Z. Wang and R. Hu are with the National Engineering Research Center for Multimedia Software, School of Computer Science, Wuhan University, Wuhan 430072, China. (e-mail: jingfang@whu.edu.cn; wzy$\_$hope@163.com; hrm@whu.edu.cn;).}
\thanks{Y. Yu is with the School of Cybersecurity, Northwestern Polytechnical University, Xi'an 710072, China (e-mail: yinboyu@nwpu.edu.cn).}
\thanks{X. Ding is with the School of Computer and Data Science, NingboTech University, Ningbo 315199, China (e-mail: XDing07@163.com).}
}

\begin{document}

\maketitle

\begin{abstract}
Image super-resolution (SR) is a technique to recover lost high-frequency information in low-resolution (LR) images. Spatial-domain information has been widely exploited to implement image SR, so a new trend is to involve frequency-domain information in SR tasks. Besides, image SR is typically application-oriented and various computer vision tasks call for image arbitrary magnification. Therefore, in this paper, we study image features in the frequency domain to design a novel scale-arbitrary image SR network. First, we statistically analyze LR-HR image pairs of several datasets under different scale factors and find that the high-frequency spectra of different images under different scale factors suffer from different degrees of degradation, but the valid low-frequency spectra tend to be retained within a certain distribution range. Then, based on this finding, we devise an adaptive scale-aware feature division mechanism using deep reinforcement learning, which can accurately and adaptively divide the frequency spectrum into the low-frequency part to be retained and the high-frequency one to be recovered. Finally, we design a scale-aware feature recovery module to capture and fuse multi-level features for reconstructing the high-frequency spectrum at arbitrary scale factors. Extensive experiments on public datasets show the superiority of our method compared with state-of-the-art methods.

\end{abstract}

\section{Introduction}

Image super-resolution (SR) is a promising approach that can recover a high-resolution (HR) image from its low-resolution (LR) counterpart. Image SR has been applied in various computer vision tasks, such as video compression \cite{ma2019image, yang2022learned}, image classification \cite{sun2019low, xie2021super}, and object detection \cite{zhou2021ecffnet, haris2021task}.
Using image spatial features is the most popular way to implement image SR, e.g., CNN-based \cite{fang2022hybrid, liu2021mdcn, zhu2022lightweight} and GAN-based networks \cite{jiang2019gan, you2022fine, chen2022learning}, which have achieved superb results. In essence, compared to HR images, LR images lack high-frequency (HF) information caused by image degradation. This difference between LR and HR images provides another significant feature (\textit{i.e.,} frequency-domain feature) for image SR \cite{li2018frequency, Yi2020singleSR, jing2020ran, cai2021freqnet, xu2022dct}. Besides, most existing SR methods focus on fixed scale factors, but arbitrary-scale SR has attracted increasing interest in many real-world applications such as image editing \cite{Son_2021_CVPR, zhu2021arbitrary} and object detection \cite{zou2020arbitrary, yun2022single}. Hence, in this paper, we aim to study arbitrary-scale image SR using frequency-domain information which remains to be an unsolved issue.

The spatial-domain image features have been exploited fully in existing SR networks \cite{zhang2021two, wu2020multi, liang2022details, zuo2020mfr}, which show compelling state-of-the-art (SOTA) performance. In recent, some researchers try to refresh the SOTA performance of SR tasks in the frequency domain \cite{ 2018Multi, Fuoli_2021_ICCV, zhao2022discrete}, such as Discrete Wavelet Transform (DWT) , Discrete Fourier Transform (DFT), and Discrete Cosine Transform (DCT). DWT has been explored in traditional model-based image SR, and recently in deep learning networks, yielding significant improvements \cite{2018Multi, Guo2017dwt}. DWT includes not only the frequency domain component of the image but also its spatial domain component. In DFT, the conversion result of the input image contains real numbers and complex numbers, which increases the computational overhead \cite{Fuoli_2021_ICCV, xue2020faster}. DCT only deals with real numbers and is more widely used \cite{zhao2022discrete, guo2019adaptivet}. These methods utilize frequency-domain information to achieve image SR from multiple perspectives and achieve excellent results.

Considering the low computational complexity of DCT, we focus on image SR using DCT domain information. Since image degradation mainly occurs on HF spectral part of an LR image, how to divide its spectrum into LF part to be retained and HF part to be recovered has not been well studied previously. A simple solution to divide the spectrum is to calculate an average threshold from a set of LR-HR paired images statistically, like \cite{guo2019adaptivet}. However, through our statistical analysis (see Section~\ref{sec:method}), we find that in the DCT domain, the spectral degradation of different images starts from different frequency points. Using a fixed threshold to divide the spectrum into LF and HF spectrum can limit the accuracy of image SR. Besides, there is a need to express arbitrary-scale image SR within a single model in real-world application scenarios \cite{zhu2021arbitrary, yun2022single}. Our statistical analysis results show that different scale factors also result in different degrees of image spectral degradation. Hence, to achieve a practical arbitrary-scale image SR approach using DCT information, it is necessary to divide image spectrum adaptively both according to scale factors and image features.


To address the above problem, we propose an end-to-end trainable \textsc{\textbf{fre}}qu\textsc{\textbf{e}}ncy domain scale-arbitrary image \textsc{\textbf{s}}uper-\textsc{\textbf{r}}esolution network, called FreeSR. We statistically analyze the spectral degradation degrees of different images at arbitrary scale factors in datasets DIV2K and Set5. Our analysis results show that the HF information of different images under different scale factors suffers from different degrees of degradation, but the LF spectra tend to be retained within a certain distribution range. Based on these results, we design an adaptive scale-aware feature division (SFD) mechanism based on deep reinforcement learning to realize a precise LF and HF spectra division for arbitrary-scale image SR tasks. Given the separated HF spectrum part of an LR image, we propose a scale-aware feature recovery (SFR) module to capture multi-level frequency features to adapt feature recovery for arbitrary-scale factors. To our best knowledge, this is the first work that implements an end-to-end trainable image arbitrary-scale SR network using full frequency-domain information. Our main contributions are summarized as follows:
\begin{itemize}
   \item We conduct a statistical analysis on different image SR datasets to learn the law of image degradation.
   \item We propose a scale-arbitrary image SR network both regarding to LF-HF division and feature adaptions in the frequency domain.
   \item Extensive experiments on public datasets (Set5, Set14, and Urban100) demonstrate the effectiveness of our SR network.
\end{itemize}


The outline of the paper is as follows. Section \ref{sec:related} reviews the related works. In Section \ref{sec:method}, we present the overview of our proposed method. The network architecture is illustrated in Section \ref{sec:network}. Experimental results and analysis are given in Section \ref{sec:experiments}. Section \ref{sec:conclusion} concludes this paper.

\section{Related Work}
\label{sec:related}
Image SR has been well-studied and achieved impressive performance due to the development of deep learning. The deep learning-based SR methods can be roughly classified into two categories according to the image processing domain: spatial-domain SR and frequency-domain SR. The existing mainstream SR methods focus on processing image SR in the spatial domain. Liu \textit{et al.}~\cite{liu2021cross} designed a cross convolution to extract edge features and achieved excellent results with more accurate edge restoration. Zhu \textit{et al.}~\cite{zhu2022lightweight} proposed a lightweight single-image SR network based on an HR-size expectation-maximization attention mechanism for better balancing performance and applicability. Hong \textit{et al.}~\cite{Hong2022WACV} utilized the different distribution characteristics of channels in SR networks, and proposed a channel-wise distribution-aware quantization scheme to improve the network efficiency. More comprehensive reviews on spatial-domain image SR are presented in~\cite{yang2019deep, lepcha2022image}. In all relevant studies, we focus on frequency-domain SR and arbitrary magnification methods, which are described in detail below.

\subsection{Frequency-domain Super Resolution}

Numerous studies have demonstrated that frequency information of HR images can be expressed through CNNs. They convert images into various frequency domains, e.g., DFT, DCT, and DWT. Li \textit{et al.} \cite{li2018frequency} transformed image super-resolution network into the frequency domain relying on convolution theorem. They cast convolutions in the spatial domain as products in the frequency domain, and non-linearity in the spatial domain as convolutions in the frequency domain, which is computationally efficient. Xue \textit{et al.}~\cite{xue2020faster} proposed a compact frequency domain neural network to reduce the model size of FNNSR. By using multiple convolutional layers with activation to learn the underlying structure, IFNNSR improved the quality of reconstructed images in the frequency domain. Guo \textit{et al.} \cite{guo2019adaptivet} integrated DCT into the network as a convolutional DCT (CDCT) layer and conducted a DCT deep SR network (ORDSR), achieving the best performance at the time. Xu \textit{et al.} \cite{xu2022dct} innovatively designed a DCT spatial cube to achieve multi-level feature decomposition of LR images, and then revised an adaptive non-local double attention mechanism to achieve HF feature extraction adaptively, which can recover the HF information of the images effectively. Unlike the above two transforms, DWT is concerned with local spatial-domain information as well as local frequency-domain information. Alireza \textit{et al.} \cite{Alireza2021} proposed a lightweight multi-domain SR network (SRNSSI), which is designed with a multi-domain residual block including a spatial-domain feature processing module for learning spatial information and a frequency-domain feature processing module for learning spectral information, enhancing the performance of the lightweight SR network. Xin \textit{et al.} \cite{Xin2022wave} decomposed the LR image into a series of wavelet coefficients (WCs), constructed a WCs prediction model to achieve the efficient and accurate reconstruction of the corresponding series of HR WCs, and then reconstructed the HR image. These methods focus on certain integer scale factors (e.g., X2, X3, X4), and each scale factor corresponds to a specific magnification module, which is thus not practical in real-world scenarios.

\begin{figure}[tb!]
    \centering
    \includegraphics[width=0.9\columnwidth]{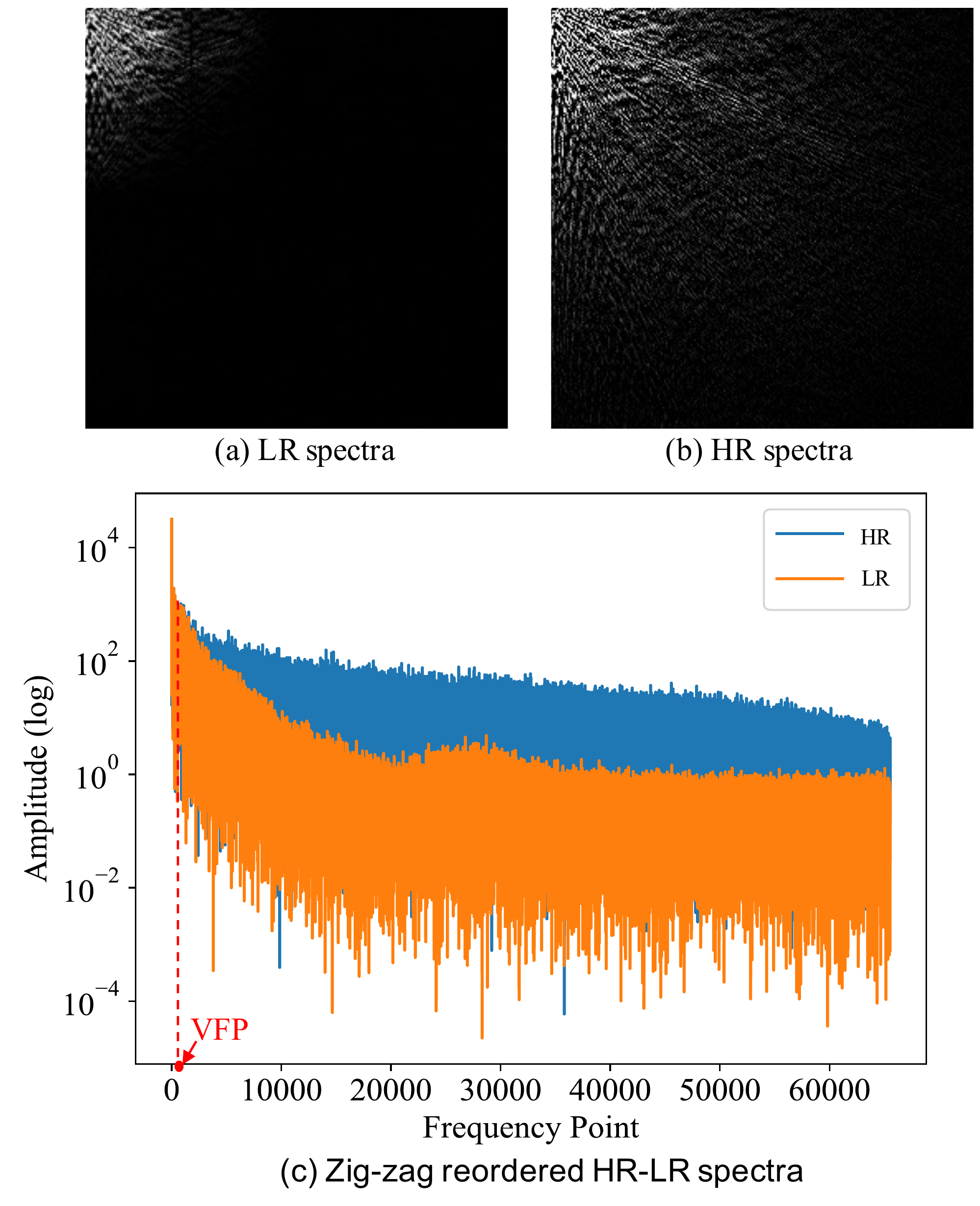}
    \caption{(a) The spectra of bicubic-interpolation up-sampled LR image ``butterfly''. (b) The spectra of its corresponding HR image. (c) The zig-zag reordered HR-LR spectra.}
    \label{fig:VFP}
\end{figure}

\begin{figure*}[tb!]
    \centering
    \includegraphics[width=0.9\textwidth]{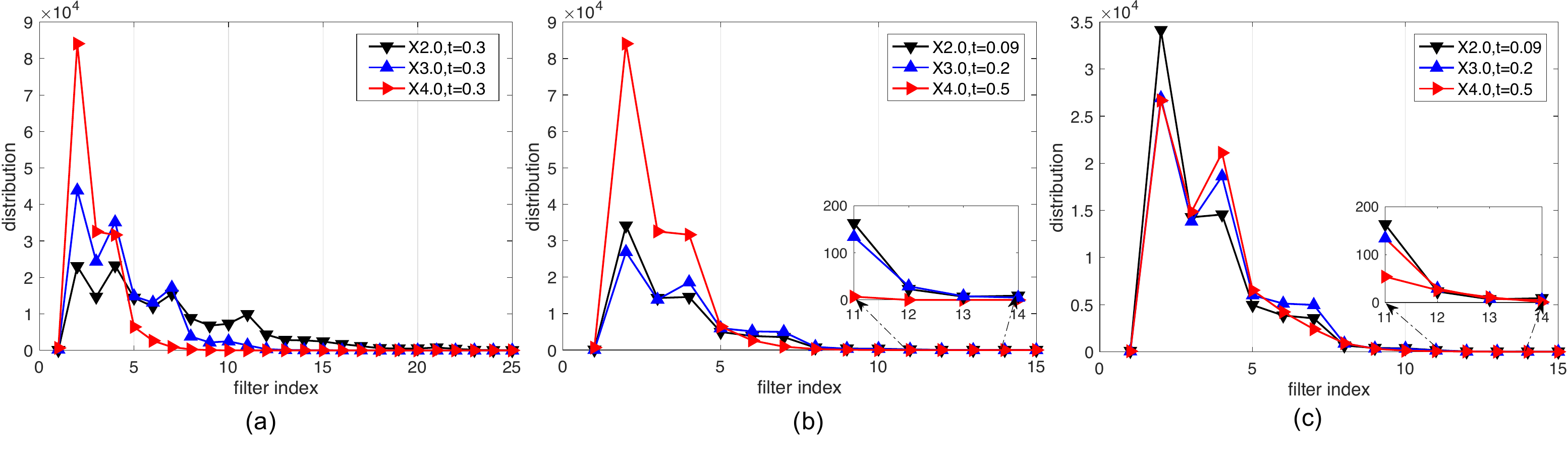}
    \caption{Statistical analysis of the frequency features of HR-LR image pairs in dataset Set14 and DIV2K.}
    \label{fig:analysis}
\end{figure*}

\subsection{Arbitrary-Scale Super Resolution}
In the past few years, deep learning-based image super-resolution has been widely studied, and these methods typically train a model for a fixed scale factor. Wang \textit{et al.}~\cite{wang2018resolution} trained a single SR model for multiple fixed integer scale factors using a resolution-aware network (RAN) which consists of two sub-networks: a decision sub-network and an upsampling sub-network.
Hu \textit{et al.} \cite{hu2019meta} were the first to propose an arbitrary magnification network with only one model (Meta-SR). In most existing fixed-scale amplification networks, the parameters of the filters are learned directly from the training dataset. Unlike these methods, Meta-SR is designed with a weight prediction layer that can be trained to predict the weights for arbitrary-scale factors. Wang \textit{et al.}~\cite{wang2021learning} proposed a plug-in module that consists of a scale-aware up-sampling layer and multiple scale-aware feature adaption blocks, called ArbSR. The plug-in module enables the existing SR models to implement arbitrary magnification, where scale factors along horizontal and vertical directions could be different. Behjati \textit{et al.}~\cite{behjati2021overnet} designed a recursive structure of skip and dense connections to obtain lightweight feature extraction and realize single image arbitrary-scale super-resolution with a single model (OverNet). Pan \textit{et al.} \cite{pan2022towards} regarded arbitrary upscaling and downscaling as one unified process and proposed a bidirectional arbitrary image rescaling network (BAIRNet) which using joint optimizing of arbitrary upscaling and downscaling to guarantee both upscaling accuracy and downscaling perception-quality. Yun \textit{et al.} proposed an arbitrary magnification SR network ($H^{2}A^{2}SR$), which performs feature adaption in the spatial domain, but magnificates features using zero padding and extracts HF components with a heurstic mask in the DCT domain. With an attention netowrk, $H^{2}A^{2}SR$ has a substantial impact on SR image quality.
As for real-world scenarios, Zhu \textit{et al.}~\cite{zhu2021arbitrary} coupled meta-learning with GAN to realize medical image arbitrary magnification. Fang \textit{et al.}~\cite{9861276} analyzed the edge explicit expression of satellite images and proposed an arbitrary-scale super-resolution network for satellite imagery based on edge enhancement. Different from these approaches, we use full frequency domain information to design an end-to-end trainable network for arbitrary-scale image SR.

\section{Overview}
\label{sec:method}

\subsection{Motivation}
\label{subsec:moti}

LR images are typically output by sensors with limited perception capabilities or down-sampled from HR images for saving storage space. Compared to HR images, LR images suffer from high information degradation in terms of high frequency. Fig. \ref{fig:VFP}(a) and (b) are examples of LR and HR images (``butterfly'' in Set5 \cite{bevilacqua2012low}) in the DCT domain. Note that the LR image is magnified by 4.0 using bicubic interpolation so that it has the same scale as the HR image. In the DCT image, the LF information of the original image is distributed in the top left, and HF information is distributed in the down right. We can use the zig-zag algorithm to roll the image from the top left and can obtain the DCT spectra from low to high frequency points. Fig. \ref{fig:VFP}(c) illustrates the degradation degrees between the LR and HR images along the frequency points. We can find that within a small range of LF points, the LR image shares the frequency information similar to the HR image. We call the end frequency point of the range as \textit{a valid frequency point} (VFP). After the VFP, the LR spectra have significant information degradation. VFP can split the image DCT spectrum into LF information and HF information. Hence, besides being magnified to a target scale, the essence of image SR for an LR image is to restore that part of the lost HF information after the VFP. However, due to complex sources of LR images, how to find the VFP for different images is still challenging. To this end, this paper presents a statistical analysis of the image degradation degree to provide prior knowledge for deciding the value of VFP and restoring image HF information.

 To establish a reasonable VFP search method, we statistically analyze several classical SR datasets (including DIV2K and Set14) to compare the differences in the degradation degree of different LR images at different scales. We randomly select 8$\times$8 blocks in each HR-LR image pair of these datasets and compare their DCT transformed frequency-domain features $f_{HR}$ and $f_{LR}$. To analyze the difference between the frequency-domain features of HR and LR images, we normalize the difference between $f_{HR}$ and $f_{LR}$ as shown in the following equation:
\begin{equation}
  f_{d} =|f_{HR}-f_{LR}|/f_{HR}.
\end{equation}

\begin{figure*}[tb!]
    \centering
    \includegraphics[width=0.9\textwidth]{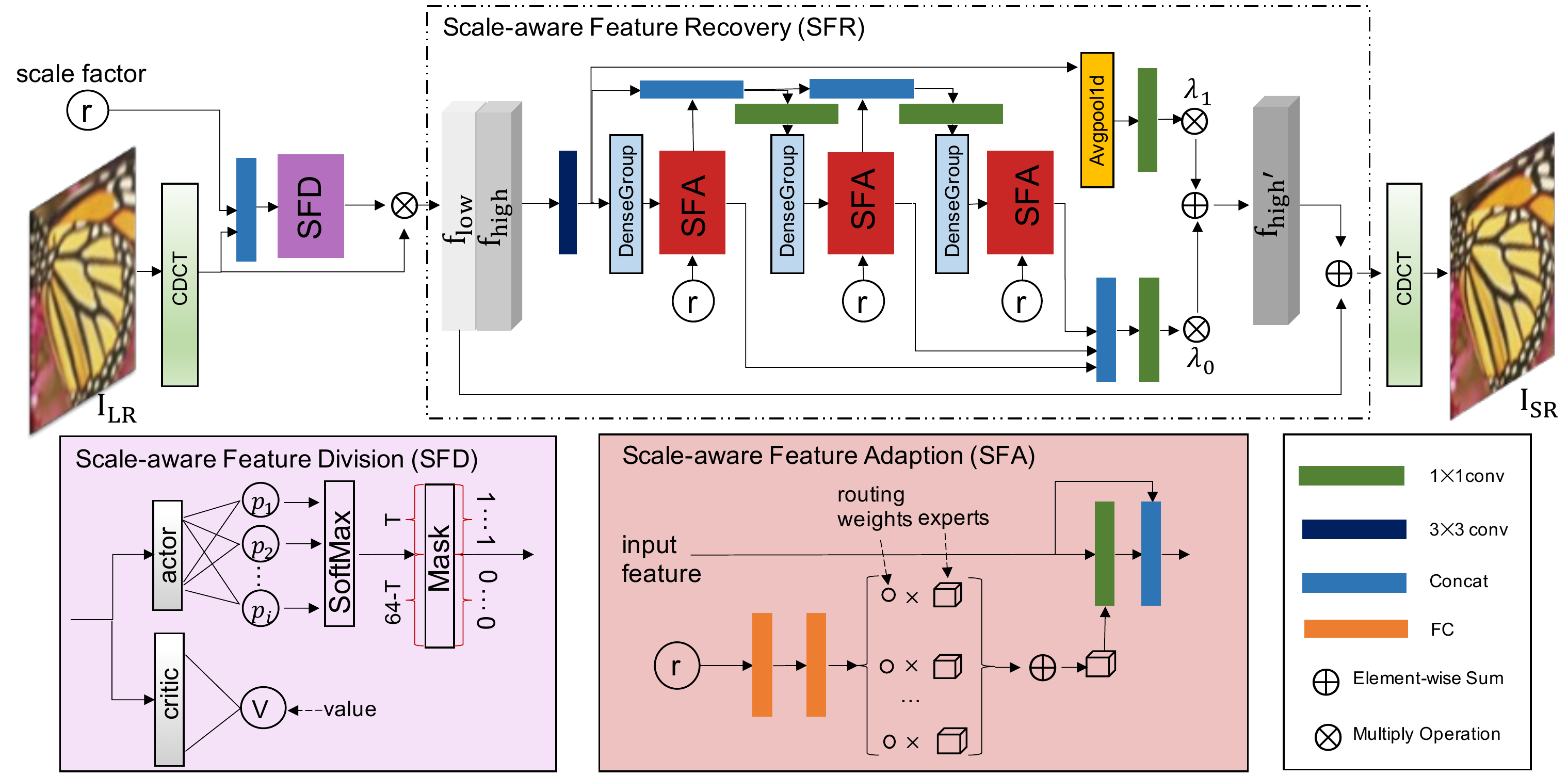}
    \caption{The general framework of FreeSR. The input of network is the bicubic enlarged LR image $I_{LR}$ and the scale factor r, and the output of the network is the SR result in spatial domain.}
    \label{fig:architecture}
\end{figure*}

We set a threshold T to obtain the VFP from the normalized difference $f_{d}$. A VFP $F$ is a frequency point, denoted as $\forall i\in[0, F], f_d(i)<T$. For example, we use $T=0.3$ and show the statistical distribution of VFPs for the Set14 dataset in Fig. \ref{fig:analysis}(a). We can find that at the same magnification scale, VFP has different values, rather than a single one, and between different scales, the values of VFP have different distributions. For example, at scale factor X2.0, the distribution of VFP ranges from 1 to 18, but at scale factor X4.0, the distribution only ranges from 1 to 7. Hence, the degradation degrees at different scales, or in different regions at the same scale have significant differences.

Considering that LR degradation is more severe as the scale factor increases, we should have a higher tolerance of the error between LR and HR images when the scale factor is larger. That is the threshold $T$ for identifying the VFP should be different at different scales. Hence, after our repeated tests, we set the experience thresholds, $T$, as 0.09, 0.2, and 0.5 for scale factors X2.0, X3.0, and X4.0, respectively. Fig. \ref{fig:analysis}(b) illustrates the VFP distribution of over seventy thousand $8\times 8$ blocks from the Set14 dataset. It can be seen that the maximum of VFPs at three scale factors are all not more than 13. Most of the VFPs are located in the range of $[2,4]$ (81.7\% at X2.0, 76.8\% at X3.0, and 81.3\% at X4.0). Similarly, we also analyze the DIV2K dataset with these same thresholds and show the distributions of VFP in Fig. \ref{fig:analysis}(c). We can also find that in DIV2K, the maximum value of VFP for all three scale factors is not more than 13. In summary, the above results suggest that we should select different VFPs in the range of 1 to 13 according to different image features and scale factors when performing image SR.

\subsection{Scale-aware Arbitrary Magnification Network}

Our goal is to realize image arbitrary-scale SR using frequency (DCT) domain information. According to the above findings, the arbitrary-scale SR method should be scale-aware both in dividing low and high frequency points and restoring the lost HF information. To this end, we design an end-to-end trainable scale-aware SR network architecture (shown in Fig. \ref{fig:architecture}). As shown in Fig. \ref{fig:analysis}(a), the degradation of HF information varies for different scale factors, so when performing image arbitrary SR, our network considers two challenges: the first one is the adaptive identification of VFP; the second one is the scale-aware high-frequency feature recovery.

\begin{itemize}
\item To address the first challenge, we design a threshold selection mechanism based on deep reinforcement learning (DRL). By extracting the current LR's frequency-domain features and the scale factor as the state input, our mechanism can automatically generate a reasonable VFP. We use the MSE loss of the final SR and HR as the reward function of DRL to optimize the DRL policy.
\item For the second challenge, we design a scale-aware feature recovery module in the DCT domain to capture multi-level features at arbitrary-scale factors. By utilizing a recursive structure, we improve the learning ability of the network and extract rich HF components.
\end{itemize}

\section{Methodology}
In this section, we present our proposed FreeSR in detail.
\label{sec:network}

\subsection{Frequency Domain Transformation}
As a special Fourier transform, discrete cosine transform (DCT) is lossless and reversible. Moreover, both its input and output are real numbers \cite{khayam2003discrete}, DCT does not involve the calculation of the imaginary part, so it can reduce the computational complexity of networks compared to Fourier transformation. Assume that image $I(x, y)$ of size $H\times W$ in the spatial domain can be divided into $H/N \times W/N $ blocks of size $N\times N$, for the $(i, j)^{th}$ block, the DCT coefficients are described as:
\begin{equation}\label{equ:DCT}
C_{i,j}(u,v)=\sum_{x=0}^{N-1}\sum_{y=0}^{N-1}I_{i,j}(x,y)\times w_{u,v}(x,y),
\end{equation}
\noindent where $ u, v \in \left \{ 0,1,2,...,N-1 \right \}$, and $w_{u, v}(x, y)$ is the DCT basis function, defined as:
\begin{equation}\label{equ:w}
w_{u,v}(x,y)=C_{u,v}cos[\frac{\pi }{N}(x+\frac{1}{2})u]\times cos[\frac{\pi }{N}(y+\frac{1}{2})v],
\end{equation}
\begin{equation}\label{equ:c}
C_{u,v}=\frac{\sqrt{1+\alpha (u)}\sqrt{1+\alpha (v)}}{N},
\end{equation}
\begin{equation}\label{equ:u}
\alpha(u)=\left\{\begin{matrix}
1 &if &u=0,   \\
0 &otherwise .   \\
\end{matrix}\right.
\end{equation}
\begin{equation}\label{equ:au}
\alpha (v)=\left\{\begin{matrix}
1 &if &v=0, \\
0 &otherwise. \\
\end{matrix}\right.
\end{equation}

Through a zigzag reorder, the DCT basis $\left\{ w_{u,v} \right\}_{u,v=1,1}^{N,N}$ can be represented as $\left\{ w_{i} \right\}_{i=1}^{N \times N}$. To establish an end-to-end network, in our method, the LR images $I_{LR}$ are transformed into the frequency domain by a trainable CDCT layer \cite{guo2019adaptivet}. We set N=8, thus, there are 64 filters $\left\{ w_{i} \right\}_{i=1}^{64}$ of size 8 $\times$ 8 in the CDCT layer. The input image x is convoluted with $\left\{ w_{i} \right\}_{i=1}^{64}$ to produce 64 frequency maps $\left\{ f_{i} \right\}_{i=1}^{64}$ of the entire image with a stride of S. Eq. \ref{equ:feature map} performs a convolution of an input image with a CDCT layer.
\begin{equation}\label{equ:feature map}
f_{i}=w_{i}\ast x, \forall i\in \left\{ 1,...,64 \right\}.
\end{equation}

With the above step, we obtain the spectral map $\left\{ f_{i} \right\}_{i=1}^{64}$ of the LR image $I_{LR}$.

\begin{figure*}[tb!]
    \centering
    \includegraphics[width=1\textwidth]{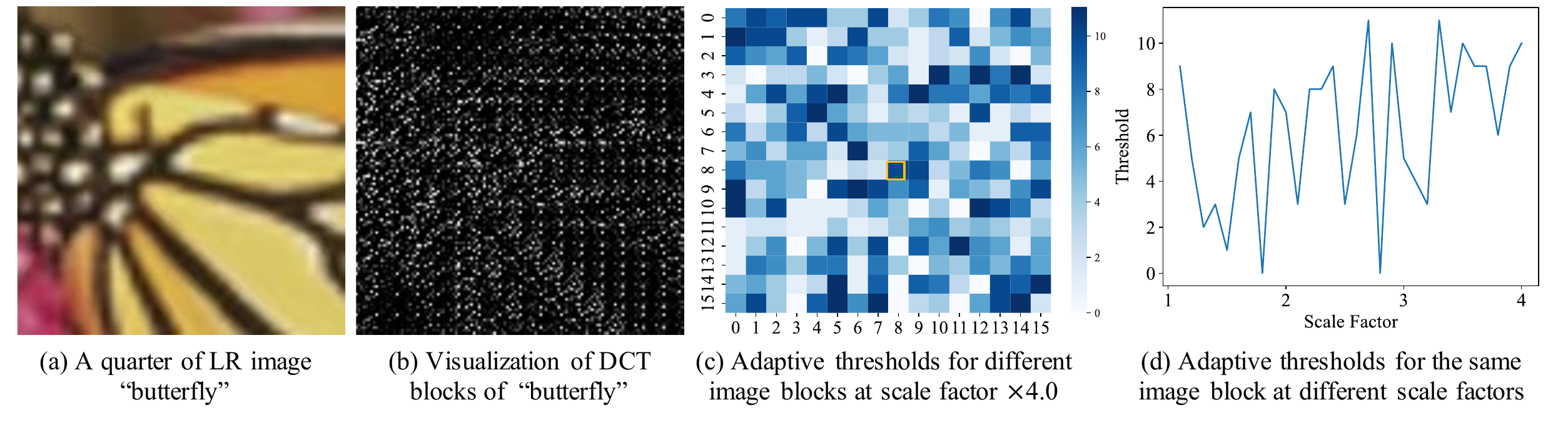}
    \caption{The visualization of proposed adaptive VFP selection mechanism.}
    \label{fig:visual}
\end{figure*}

\subsection{Scale-aware Feature Division}

Given a DCT spectral map $f$ of the bicubic-upsampled LR image, we need to identify the VFP $F$ for $f$ and divide $f$ into a HF spectral map $f_{high}$ and a LF spectral map $f_{low}$. As analyzed in Sec. \ref{subsec:moti}, the VFP $F$ varies with scale factors and image contents. It is necessary to select different VFPs for different scale factors and different image blocks. The selection of VFP is an unsupervised learning task, and the goodness of VFP selection needs to be evaluated by the final image SR performance. Moreover, the value of VFP can also affect the effectiveness of HF information reconstruction and image SR, the degree of difference between the final SR image and the HR image can be used to motivate the selection of VFP. Therefore, we introduce reinforcement learning to optimize the VFP selection process by using the MSE loss of SR images and HR images as a reward value. Meanwhile, in order to build an end-to-end image SR deep learning model, we use a deep reinforcement learning network to implement adaptive VFP selection, thus realizing an end-to-end deep image SR network from image DCT transformation, LF-HF feature division, and HF feature reconstruction, ensuring the cooperative optimization of parameters in each module. Hence, based on on-policy actor-critic DRL, we design a scale-aware feature division (SFD) block in our FreeSR network to accomplish an adaptive LF-HF feature division.


The task of finding the precise VFP $F$ for an image block can be formulated as a Markov decision process (MDP). Formally, an MDP can be described as a 5-tuple $\left \langle S, A, P, R, \gamma \right \rangle $, where $S$, $A$, $R$, and $\gamma$ are the state space, action space, rewards, and discount factor, respectively. DRL is typically used to address sequential decision-making problems. But, the task of finding VFP $F$ is an instant decision problem. $\gamma\in [0,1]$ specifies the importance between future rewards and the current reward, \textit{i.e.}, $\gamma=0$ represents an agent only concerned with its immediate reward, while $\gamma=1$ denotes an agent striving for a long-term higher reward. Hence, by setting $\gamma=0$, we perform the MDP in one step for the task of finding VFP. We describe the MDP in detail as follows:


\textbf{State Space}: In order to realize scale-aware property, we cascade the LR image (8$\times$8) spectra (64 dimensions) and a scale factor $r$ to formulate the state $s$ (a 65-dimension array):
\begin{equation}
    s=[f\quad r].
\end{equation}

\textbf{Action Space}: An action $a\in A$ is the VFP for spectra division. As analyzed in Sec.\ref{subsec:moti}, the largest VFP for different DCT spectral maps (64 dimensions) at arbitrary-scale factors is 13. Hence, we set the action space $A$ to be $[1,13]$. Given an action $a\in [1,13]$, we generate a 64-dimension mask $M$, in which the value whose index is less than $a$ is equal to be 1, otherwise 0:
\begin{equation}
    M = [\underbrace{1\cdots 1}_{a}\ \underbrace{0\cdots 0}_{64-a}].
\end{equation}

\noindent With the mask $M$, we can generate a low $f_{low}$ and high $f_{high}$ spectral map by a multiply operation with the LR spectra $f_{LR}$ and $M$:
\begin{equation}\label{equ:div}
\begin{split}
f_{low} =&f_{LR}\otimes M,\\
f_{high} =&f_{LR}\otimes (1-M).
\end{split}
\end{equation}

\textbf{Reward}: The reward function is used to optimize the DRL policy. Since our goal is to make the generated SR image approach the HR counterpart, we use the MSE loss of SR and HR images to set the reward function as follows:
\begin{equation}
R=1-\left \|I_{SR} - I_{HR} \right \| _{2}^{2}.
\end{equation}

\noindent The closer the generated SR image is to the HR image, the higher the reward obtained by the DRL agent.

To integrate the optimization of the DRL policy with the major learning process for SR, we follow the actor-critic architecture to design an on-policy one-step DRL algorithm. An actor network $\pi$ is parameterized by $\theta$ and is responsible for generating actions for input states and interacting with the DRL environment. A critic network $V$ is parameterized by $\theta_v$ and is responsible for evaluating the performance of the actor and guiding action generations in the actor. By training the DRL model, we aim to obtain the policy $\pi(a|s)$ which maps each state $s$ to an action $a$ so that the selected action can maximize the expected reward. The policy can be evaluated by the state-action value function $Q_{\pi}(s, a)$, which can be expressed as follows:
\begin{equation}
\begin{split}
  Q_{\pi }(s,a)=& E_{\pi }\left [ R_{t+1} + \gamma Q_{\pi }(S_{t+1},A_{t+1})|S_{t}=s, A_{t}=a \right ]\\
   =& R_{s}^{a} + \gamma V_{\pi } (s^{'} )\\
   =& R_{s}^{a} .
  \end{split}
\end{equation}
\noindent Note that we set $\gamma=0$ for concerned with the immediate reward. Besides, $V_{\pi }(s)$ is the state value function, \textit{i.e.}, the expectation of subsequent reward values for all possible actions $A$ in the current state $s$:
\begin{figure*}[tb!]
    \centering
    \includegraphics[width=1\textwidth]{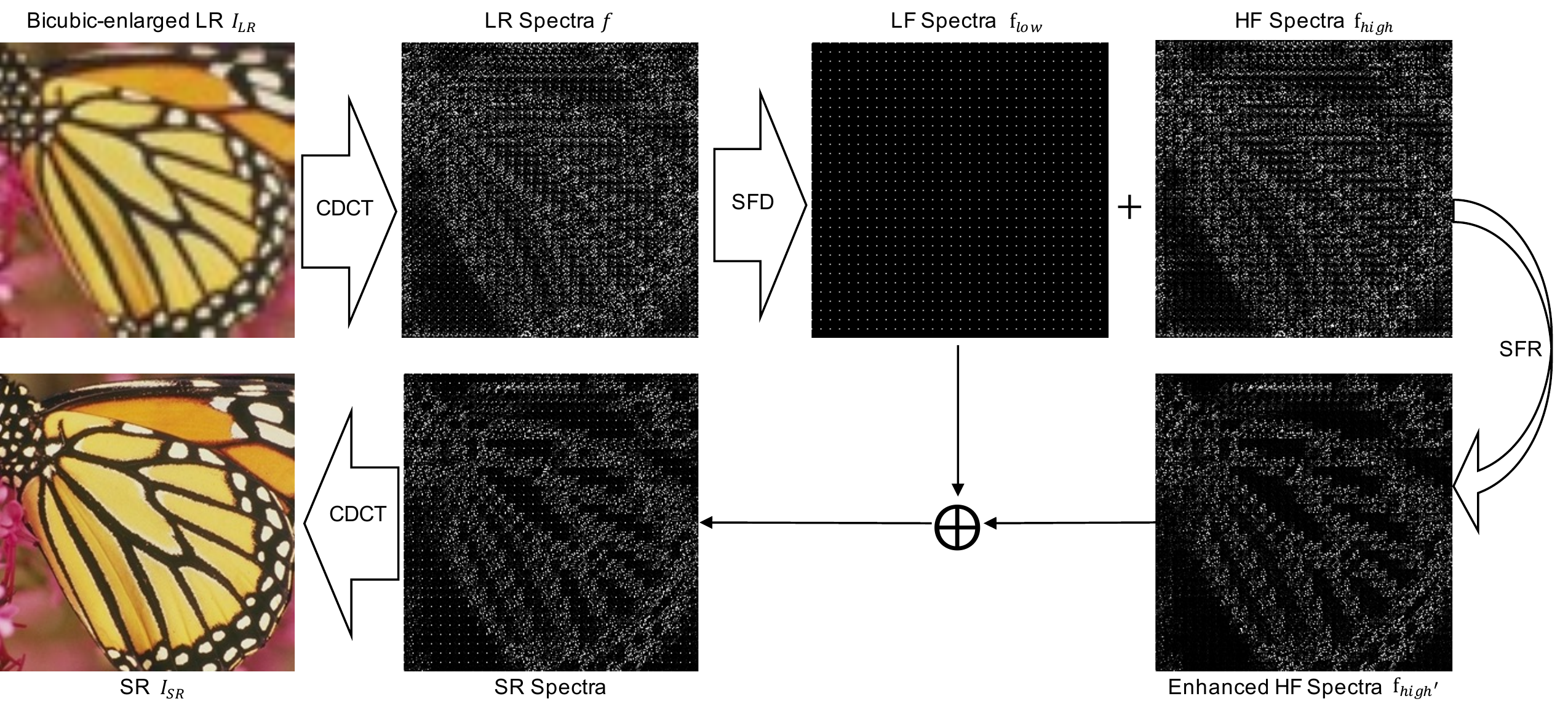}
    \caption{Schematic diagram of input and output of each network module.}
    \label{fig:sketch}
\end{figure*}
\begin{equation}
\begin{split}
V_{\pi } (s)=&E\left \{ \sum_{k=0}^{\infty }\gamma ^{k} r_{k+1}| s_{0}=s, \pi \right \}\\
=&E_{\pi }\left [ R_{t+1} \right ] .
\end{split}
\end{equation}
\noindent The goal of the actor-critic algorithm is to find the best policy possible. We use the policy gradient method to update the policy parameter $\theta$ as follows:
\begin{equation}
\label{equ:the}
    \theta := \theta + \alpha\nabla_{\theta}J(\theta),
\end{equation}
\begin{equation}
    \nabla_{\theta}J(\theta)= \frac{1}{N} \sum_{i=1}^{N} \nabla_{\theta} \log \pi_{\theta}\left(a_{i} \mid s\right) A^{\pi}\left(s, a_{i}\right),
\end{equation}
\noindent where $\alpha$ is the learning rate and $A_{\pi}$ is the advantage function and can be expressed as follows:
\begin{equation}
\begin{split}
  A_{\pi }(s,a)=& Q_{\pi }(s,a)-V_{\pi } (s)\\
  =& R_{s}^{a} - V_{\pi} (s).
  \end{split}
\end{equation}

According to the above function definition, we adopt an online training of the DRL model, i.e., when training the overall network architecture, the SR image generated by the current network and its corresponding HR image are used to calculate the action reward generated by the current DRL strategy, so as to update the DRL network parameters. This method enables effective fusion and synergistic optimization of VFP selection and HF feature recovery.

The visualization of our adaptive VFP selection mechanism is shown in Fig. \ref{fig:visual}. We display a quarter of the upsampled LR image ``butterfly'' with pixel $128\times 128$ in Fig. \ref{fig:visual}(a), and its corresponding $16\times 16$ DCT transformation blocks in Fig. \ref{fig:visual}(b), each of which stands for an $8\times 8$ image spectra. Fig. \ref{fig:visual}(c) shows the learned adaptive VFP of the $16\times 16$ blocks in Fig. \ref{fig:visual}(b) at scale factor $\times$4.0. It can be seen that different image blocks have different VFPs. Fig. \ref{fig:visual}(d) further illustrates that the VFP for the same image block highlighted in Fig. \ref{fig:visual}(c) also varies at different scale factors.

\subsection{Scale-aware Feature Recovery}

Since the main task of SR is to recover the HF information from LR images, after adaptive feature division, we further conduct a scale-aware feature recovery (SFR) module to reconstruct the HF information of images at arbitrary scale factors. The recovered HF spectra $f_{high^{'}}$ is added on $f_{low}$ to obtain the final image spectra $f_{SR}$. By a transpose convolution of the CDCT layer filters $\left\{ w_{i} \right\}_{i=1}^{64}$ and image spectra $f_{SR}$, the network outputs the final SR image $I_{SR}$.

Through the mask M generated in SFD, we obtain the HF spectra $f_{high}$ and the LF spectra $f_{low}$. The HF spectra are first sent into a $3\times 3$ convolutional layer to extract the shallow features $f_{s}$. Then, we adopt the dense group in \cite{behjati2021overnet} to achieve a lightweight feature extraction by reusing information through a recursive structure of dense and skip connections, see Figure \ref{fig:architecture}. The dense group is conducted by multiple modified versions of residual blocks (RBs) \cite{yu2018wide}, which utilize wide low-rank convolutions instead of transitional residual blocks. By exploiting the inter-dependencies between feature channels, RBs focus on more informative features. The RB process is shown as:
\begin{equation}
f_{o} =\lambda _{0} \mathit{SE}(\mathit{WA} (f_{i}))+\lambda _{i}f_{i},
\end{equation}
\noindent where $f_{i}, f_{o}$ are the input and output of a residual block, $\mathit{WA}$ is the wide activation operation, $\mathit{SE}$ is the squeeze-and-excitation operation.

The input of an RB is connected in series with the output of all previous RBs in the group and merged with a $1\times1$ convolution to form a locally dense group (LDG). The local information is progressively collected by $1\times1$ convolution.
\begin{equation}
f_{D} =conv_{1\times 1 } ([f_{0},...,f_{D-1}]),
\end{equation}
\noindent where $f_{D}$ is the output feature map of LDG, $[f_{0},...,f_{D-1}]$ is the concatenation operation of feature maps.

The output of LDG $f_{D}$ and scale factor r are then sent into our proposed Scale-aware Feature Adaption (SFA) Block to realize arbitrary-scale feature learning. The SFA block is illustrated in Figure \ref{fig:architecture}. First, the scale factor is sent into a model controller with two fully connected (FC) layers to generate routing weights. Then, we combine the routing weights with experts to achieve a scale-aware filter, in this scenario, experts stand for a set of convolutional kernels. The predicted scale-aware features and input features are adaptively merged through a $1\times 1$ convolutional layer. Finally, the merged information is cascaded with the input feature to obtain the final output feature of the SFA block.

A recursion is performed in our network to achieve lightweight feature learning in the frequency domain. To ensure that no information is lost during the reconstruction process, a long-range skip connection is added to allow access to the original information and to encourage back-propagation of the gradients. We also design a global averaging pool $``Avgpool1d''$ followed by a $1\times 1$ convolution layer to fully capture channel-wise dependencies from the aggregated information. The final HF spectra are then expressed as:
\begin{equation}
f_{high^{'}} =\lambda _{1}f_{D} + \lambda _{0}\delta (conv_{1\times 1} (GAP(conv_{1\times 1}(f_{s})))),
\end{equation}
\noindent where $f_{high^{'}}$ is the output feature map, $\delta$ is the ReLU activation, GAP is the global average pooling, $\lambda _{0}, \lambda _{1}$ are learned parameters.

Finally, to ensure the effective retention of LF information and the cascading of low- and high- frequency information, SFR uses the mask $M$ generated by the SFD module to remove the LF part of the recovered HF information $f_{high^{'}}$:
\begin{equation}
  f_{high^{'}} =f_{high^{'}}\otimes (1-M).
\end{equation}
\noindent $f_{high^{'}}$ is then added to the original $f_{low}$ to form the final spectral feature $f_{SR}$. After sending it into a CDCT layer to realize the IDCT transformation, we obtain the SR results in the spatial domain. The step-by-step results of our network are shown in Fig. \ref{fig:sketch}. It shows the low- and high- frequency characteristics and the enhanced HF spectrum in detail, including the LR image $I_{LR}$, the spectrum $f_{LR}$, the split LF spectrum $f_{low}$ and HF spectrum $f_{high}$, the reconstructed HF spectrum $f_{high^{'}}$, the final SR image spectrum $f_{SR}$, and the SR image $I_{SR}$.

\subsection{Loss Function}

\begin{figure*}[tb!]
    \centering
    \includegraphics[width=1\textwidth]{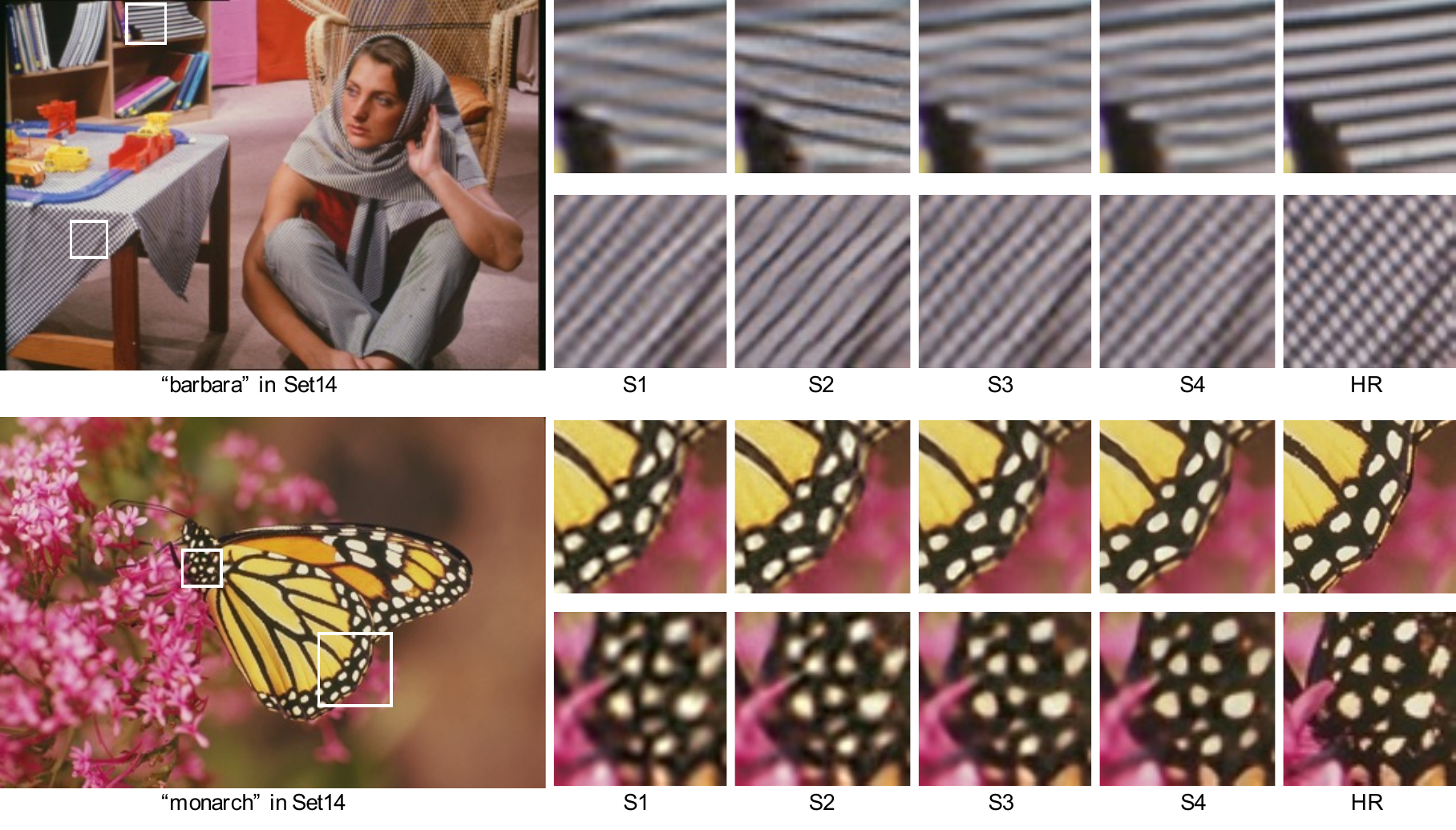}
    \caption{Qualitative ablation study of different settings on ``monarch'', ``barbara'' in Set14 under scale factor $\times$4.0. S1 is the baseline network ORDSR~\cite{guo2019adaptivet}. S2 introduces DG blocks on S1. S3 adds SFD on S2. S4 adds SFA on S3 (\textit{i.e.}, FreeSR). }
    \label{fig:abl}
\end{figure*}

According to the method design, our loss function contains three parts: DCT transformation $L_{DCT}$, actor-critic cost-to-go function $L_{SFD}$, and the frequency domain regularized loss function $L_{SFR}$. To guarantee reconstruction via the transpose convolution-based inverse, a pairwise orthogonality constraint on the filter/basis function of the CDCT layer is employed in our network.
\begin{equation}
\forall i\ne j,\left \| vec(w_{i}^{T}) vec(w_{j} ) \right \| _{2}^{2},
\end{equation}
\noindent where $i, j \in \left \{ 1,...,64 \right \} $, and $vec(\cdot ) $ is the vectorization operation that converts the matrix to a column vector. Besides, to preserve the order complexity of the DCT basis, we also introduce a regularization constraint:
\begin{equation}
\left \| var(w_{t}) - var(w_{t}^{dct} ) \right \|_{2}^{2},
\end{equation}
\noindent where $t \in \left \{ 1,...,64 \right \} $, $w_{t}$ are the filters in CDCT layer, $w_{t}^{dct}$ is the corresponding DCT basis function. $var(w)$ is the variance of a filter w given by Bessel's correction version:
\begin{equation}
var(w) = \frac{1}{N^{2}-1} \sum_{m} (w^{m}-\frac{1}{N^{2}} \sum_{n} w^{n})^{2},
\end{equation}
\noindent where $N=8$, $\sum_{m}w^{m}$ and $\sum_{n}w^{n}$ are the summation of all the elements in filter $w$. Ideally, the above two constraints should evaluate to be zero. Hence, we have the loss function $L_{DCT}$ for DCT as follows:
\begin{equation}
\begin{split}
    L_{DCT}= &\lambda \frac{1}{2} \sum _{(i,j),i\ne j }\left \| vec(w_{i}^{T}) vec(w_{j} ) \right \| _{2}^{2}  \\
    &+\mu \frac{1}{2} \sum _{t}\left \| var(w_{t}) - var(w_{t}^{dct} ) \right \|_{2}^{2},
\end{split}
\end{equation}
\noindent where $\lambda$ and $\mu$ are two weights for balancing these two constraints.

In our SFD block, the actor-critic operates in the forward, and the policy and value networks are updated at every step. According to the policy gradient method shown in Eq. (\ref{equ:the}), we use the following loss function to update the actor policy:
\begin{equation}
L_{\pi } =-\frac{1}{N} \sum_{i=0}^{N} A_{\pi_{\theta}   } (s,a)log\pi _{\theta }(s,a)+\beta H(\pi_{\theta}(s,a)),
\end{equation}
\noindent where we introduce the entropy $H$ of the policy to the objective function. Adding $H$ can improve network exploration by preventing premature convergence to suboptimal deterministic policies, and $\beta$ is a weight to achieve the modulation of the entropy of the strategy, which is experimentally verified to be optimal when set to 0.01 \cite{mnih2016asynchronous}.

The value function update usually adopts mean square error MSE as the loss function:
\begin{equation}
L_{v} = \frac{1}{2} \left \| R_s^a - V_{\pi}(s)\right \| _{2}^{2}.
\end{equation}
\noindent The total actor-critic loss in the SFD block is defined as:
\begin{equation}
L_{SFD}=\frac{1}{2}({L_{\pi }+L_{v}}).
\end{equation}

The goal of SFR is to restore the lost HF information, which is similar to feature extractions in the spatial domain. We can design a loss function for SFR, which, however, cannot constrain sufficient optimization for the end-to-end image SR. Hence, we use the MSE loss in the spatial domain to make the generated SR images approach the ground truth.
\begin{equation}
L_{SFR}=\frac{1}{2} \left \| I_{SR}-I_{HR} \right \| _{2}^{2}.
\end{equation}
\noindent Therefore, the total loss of our SR network is expressed as:
\begin{equation}
L_{total}=L_{SFR} + L_{DCT}+\omega L_{SFD},
\end{equation}
\noindent where $\omega $ is used to modulate the weight of the reinforcement learning network, which is experimentally verified to be optimal when set to 0.1.

\section{EXPERIMENTS}
\label{sec:experiments}
\subsection{Data and Settings}
We conduct experiments on the widely used Set14 dataset \cite{schulter2015fast} and DIV2K dataset \cite{agustsson2017ntire}. We crop all HR images to size 96 × 96 as training data, the HR images are down-sampled with a total of 30 scale factors in steps of 0.1 in the range of [1.1, 4.0], and then up-sampled using bicubic interpolation to form LR images with the same size as the HR images. To ensure the robustness and generalization of our method, we choose several commonly used datasets for testing: Set5 \cite{bevilacqua2012low}, Set14 \cite{schulter2015fast}, Urban100 \cite{2015Single}, and DIV2K \cite{agustsson2017ntire}. Our experiments are conducted on an NVIDIA GTX 3090Ti GPU with the PyTorch platform. The learning rate is initialized to $2\times 10^{-4}$, and we add CosineAnnealingLR to change the learning rate following \cite{mishra2019improving}.

Two commonly used full-reference image quality evaluation metrics: peak signal-to-noise ratio (PSNR) and structural similarity (SSIM), are introduced in our work to evaluate the performance of different SR methods. Moreover, in real-world scenarios, we only have LR images without corresponding HR reference images. Hence, the no-reference image quality evaluation indices such as natural image quality evaluator (NIQE) \cite{mittal2012making} and spatial-spectral entropy-based quality (SSEQ) \cite{liu2014no} are introduced for further quality evaluation. Note that, all metrics were calculated in the luminance channel.

\subsection{Ablation Study}

To assess the effectiveness of our proposed SFD and SFR, we conduct an ablation study on the Set14 dataset. Since the trainable CDCT layer is proposed in ORDSR \cite{guo2019adaptivet} which is an image SR network in the frequency domain, we use it as the baseline. Hence, we have the first setting \textbf{S1} the same as the ORDSR network, which has a CDCT layer for transforming an image into the DCT domain, a fixed threshold (3 for $\times$4.0) to divide LF and HF spectra, and multi-layer CNNs (15 layers) as the feature learning module. Based on \textbf{S1}, we further analyze the impact of the dense group (DG) used in our network (\textbf{S2}), in which we replace the 15 CNN layers with 3 DGs with the same connections as our SFR. Then, we include our SFD module into \textbf{S2} as the setting \textbf{S3} to evaluate the necessity of adaptive VFP selection. The last setting \textbf{S4} applies both our SFD and SFR to investigate the overall effect of our proposed frequency-domain scale-aware arbitrary image SR (FreeSR).

\begin{figure}[tb!]
    \centering
    \includegraphics[width=0.9\columnwidth]{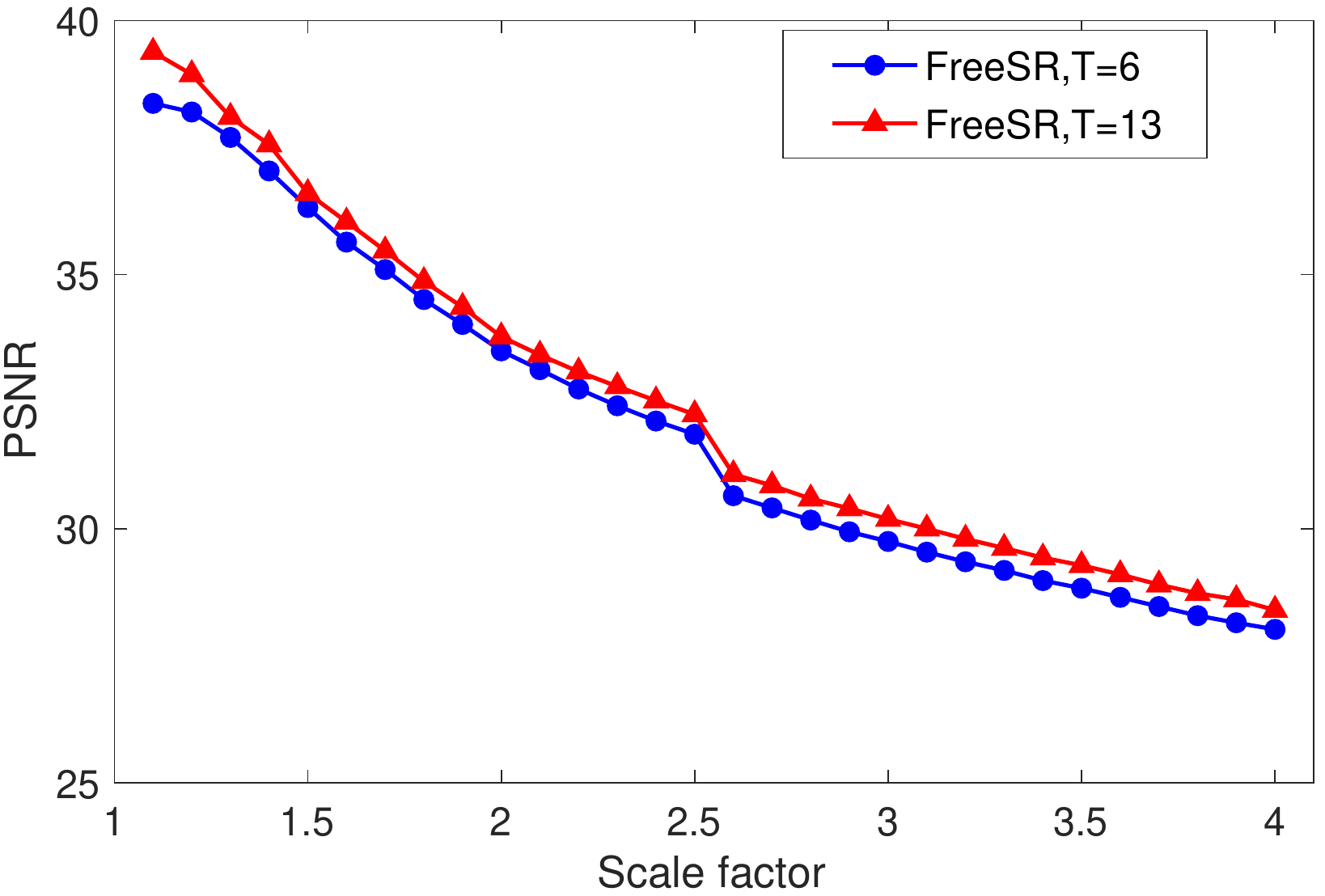}
    \caption{Ablation study of VFP of scale-aware feature division module.}
    \label{fig:tcolor}
\end{figure}

\begin{table}[]
\centering
\caption{PSNR/SSIM results achieved by our network with different settings on Set14 of scale factor $\times$4.0.}
\label{tab:abl}
\scalebox{1}{
\begin{tabular}{|l|l|l|l|l|l|l|l|}
\hline
Model & \multirow{2}{*}{SFD} & \multicolumn{2}{c|}{SFR}  & \multirow{2}{*}{PSNR$\uparrow$/SSIM$\uparrow$} \\ \cline{1-1} \cline{3-4}
Settings &   & SFA  &DG    &   \\ \hline
S1        & \ding{55} & \ding{55} & \ding{55}   & 28.33/0.776 \\ \hline
S2        & \ding{55} & \ding{55} &\ding{51}    & 28.57/0.781 \\ \hline
S3        & \ding{51} & \ding{55} &\ding{51}    & 28.89/0.790 \\ \hline
S4        & \ding{51} & \ding{51} &\ding{51}    & $\textbf{29.05/0.794} $\\ \hline
\end{tabular}
}
\end{table}

\begin{figure*}[tb!]
    \centering
    \includegraphics[width=1\textwidth]{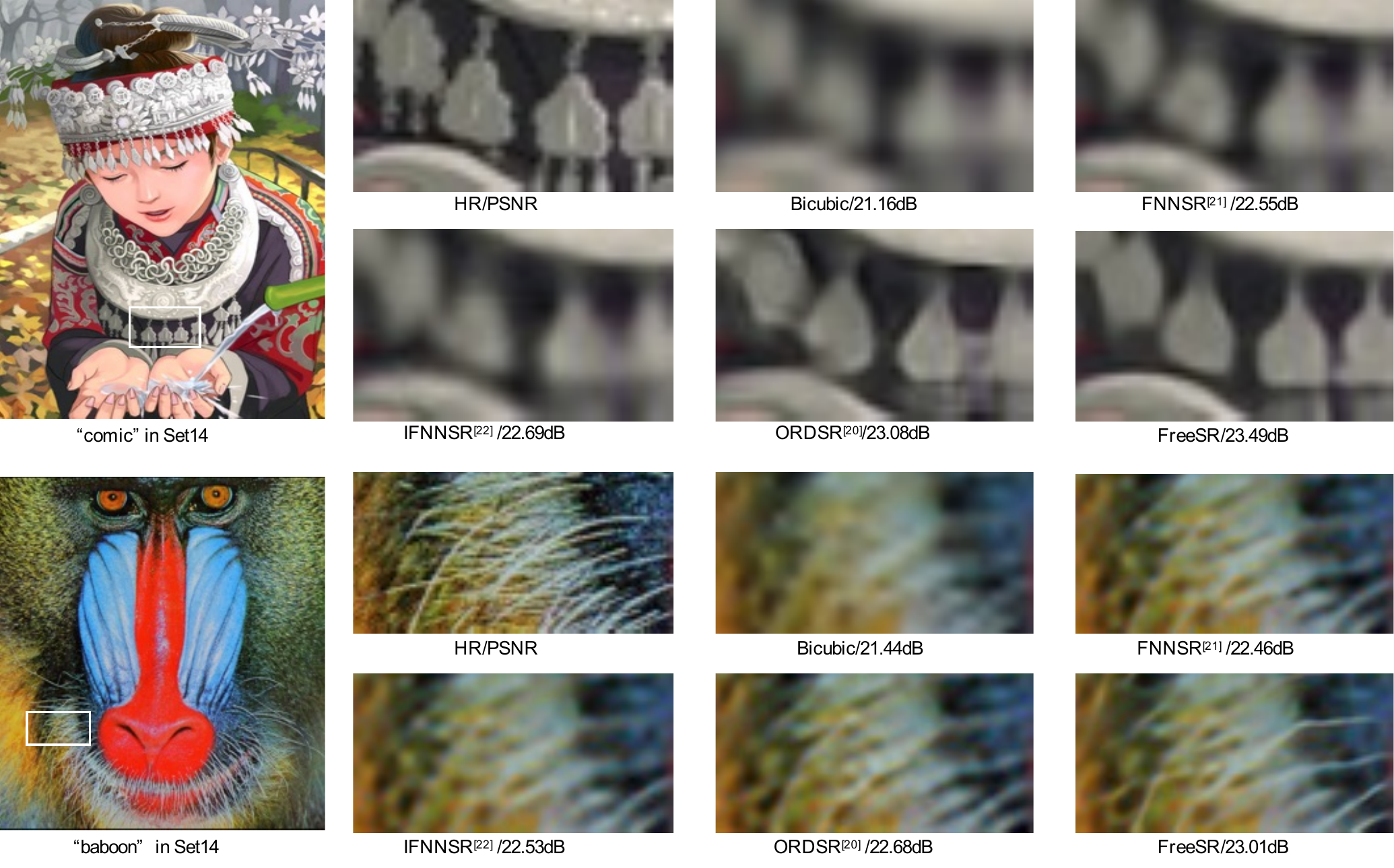}
    \caption{Recovered image of different frequency-domain methods under scale factor $\times$4.0.}
    \label{fig:visualf}
\end{figure*}

It can be observed from Table \ref{tab:abl} that the performance benefits from DG, SFD, and SFA. Compared with \textbf{S1}, the PSNR and SSIM values of \textbf{S2} are improved by 0.24dB and 0.64$\%$. That is because DG can learn richer features through dense and skip connections than multi-layer CNNs in \textbf{S1}. When adding SFD to \textbf{S2}, the average PSNR and SSIM values of \textbf{S3} are significantly improved by 0.32 dB and 1.2$\%$, respectively. This shows that adaptive segmentation of LF and HF information can achieve better performance than using fixed threshold segmentation. With our SFA blocks, the performance of \textbf{S4} is further improved by 0.16dB and 0.5$\%$ compared to \textbf{S3}. Both SFD and SFA can improve the performance of the algorithm and SFD contributes more to our network.

\begin{table}[]
\centering
\label{tab:result3}
\caption{Comparison results of PSNR$\uparrow$/SSIM$\uparrow$ on different datasets with scale factor of 2, 3, and 4. Bold indicates the best performance.}
\scalebox{0.85}{
\begin{tabular}{|c|c|cccc|}
\hline
\multirow{2}{*}{dataset}  & \multirow{2}{*}{scale} & \multicolumn{4}{c|}{PSNR$\uparrow$/SSIM$\uparrow$}                                                                                       \\ \cline{3-6}
                          &                        & \multicolumn{1}{c|}{FNNSR \cite{li2018frequency}}       & \multicolumn{1}{c|}{IFNNSR \cite{xue2020faster}}      & \multicolumn{1}{c|}{ORDSR \cite{guo2019adaptivet}}       & FreeSR        \\ \hline
\multirow{3}{*}{Set5}     & $\times$2.0                      & \multicolumn{1}{c|}{35.18/0.941} & \multicolumn{1}{c|}{35.65/0.951} & \multicolumn{1}{c|}{35.53/0.941} & \textbf{36.83/0.964} \\ \cline{2-6}
                          & $\times$3.0                      & \multicolumn{1}{c|}{31.39/0.880} & \multicolumn{1}{c|}{31.82/0.890} & \multicolumn{1}{c|}{32.22/0.890} & \textbf{34.68/0.935} \\ \cline{2-6}
                          & $\times$4.0                      & \multicolumn{1}{c|}{29.31/0.823} & \multicolumn{1}{c|}{29.76/0.840} & \multicolumn{1}{c|}{30.52/0.852} & \textbf{32.51/0.897} \\ \hline
\multirow{3}{*}{Set14}    & $\times$2.0                      & \multicolumn{1}{c|}{31.38/0.893} & \multicolumn{1}{c|}{31.74/0.903} & \multicolumn{1}{c|}{32.57/0.909} & \textbf{33.67/0.917} \\ \cline{2-6}
                          & $\times$3.0                      & \multicolumn{1}{c|}{28.29/0.799} & \multicolumn{1}{c|}{28.73/0.822} & \multicolumn{1}{c|}{29.82/0.831} & \textbf{30.85/0.85}  \\ \cline{2-6}
                          & $\times$4.0                      & \multicolumn{1}{c|}{26.58/0.723} & \multicolumn{1}{c|}{26.97/0.737} & \multicolumn{1}{c|}{28.34/0.777} & \textbf{29.05/0.794} \\ \hline
\multirow{3}{*}{Urban100} & $\times$2.0                      & \multicolumn{1}{c|}{27.86/0.864} & \multicolumn{1}{c|}{28.16/0.881} & \multicolumn{1}{c|}{29.05/0.886} & \textbf{30.70/0.911} \\ \cline{2-6}
                          & $\times$3.0                      & \multicolumn{1}{c|}{25.16/0.757} & \multicolumn{1}{c|}{25.48/0.782} & \multicolumn{1}{c|}{26.38/0.800} & \textbf{27.56/0.841} \\ \cline{2-6}
                          & $\times$4.0                      & \multicolumn{1}{c|}{24.40/0.709} & \multicolumn{1}{c|}{24.77/0.734} & \multicolumn{1}{c|}{25.01/0.740} & \textbf{25.62/0.772} \\ \hline
\end{tabular}}
\end{table}

\begin{figure*}[tb!]
    \centering
    \includegraphics[width=1\textwidth]{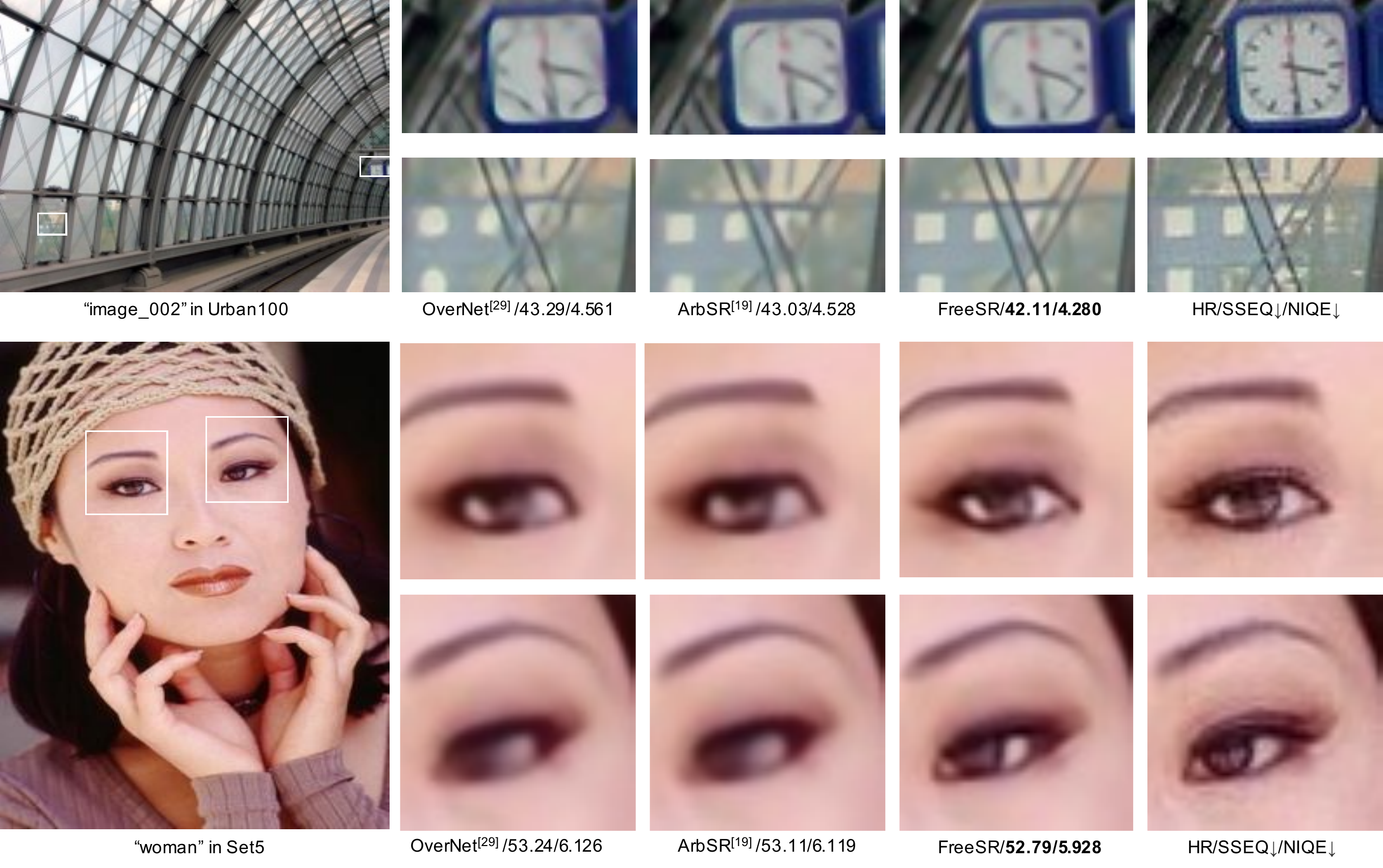}
    \caption{Recovered image of different arbitrary magnification methods under scale factor $\times$3.5.}
    \label{fig:3.5}
\end{figure*}

We further show the visual comparison results of different settings in Fig. \ref{fig:abl}. It can be observed that the image reconstruction performance gradually improves as the modules are introduced one by one. For image ``monarch'', the enlarged parts show the body and wing of the butterfly, and the reconstructed image of \textbf{S4} has clearer and finer details than other settings. For image ``barbara'', both \textbf{S1}, \textbf{S2}, \textbf{S3} generate a deformed contour of the books, and the book contour generated by \textbf{S4} is more accurate. It suggests that both SFD and SFR are indispensable in our approach. Therefore, both SFD and SFR can significantly improve the visual effect of our network.

In Section~\ref{subsec:moti}, we qualitatively analyze the VFP distributions of more than 70,000 image blocks, and the maximum VFP distributions at different scale factors are close to [1, 13], but most of them are concentrated in the range of [2, 4]. To further demonstrate that different threshold selection leads to differences in network performance, we train FreeSR under different VFP selection ranges (i.e., the action space size in SFD), and then compare the impact of different VFP selection ranges on the SR network performance. Here, we choose two different action spaces [1, 6] and [1, 13], and evaluate the PSNR of the reconstructed images under 30 scale factors. Fig. \ref{fig:tcolor} shows the final comparison results. It suggests that the FreeSR network can achieve better reconstruction performance when a larger VFP selection range is set. Notice that the larger the VFP, the slower the network convergence. To trade off network performance and training time, we set action space as [1, 13] during our training process.

\subsection{Comparison with State-of-the-art Methods}

In this section, we compare the proposed FreeSR with representative and SOTA SR methods, both fixed scale and arbitrary scale. Since our work is complemented in the frequency domain, we first compare it with representative frequency-domain SR networks, such as FNNSR \cite{li2018frequency}, IFNNSR \cite{xue2020faster}, ORDSR \cite{guo2019adaptivet}, on integer scale factors X2/X3/X4. The experiments are conducted on widely used benchmark datasets Set5, Set14, and Urban100. The PSNR/SSIM results are shown in Table \ref{tab:result3}, bold represents the best results. It can be observed from Table \ref{tab:result3} that our FreeSR achieves the best performance compared to other frequency-domain SR networks. For example, at a scale factor of 4.0, the average PSNR and SSIM of FreeSR are improved by 1.2 dB and 0.01 over ORDSR on the Set5 dataset.
These frequency-domain SR networks try to recover as much HF information as possible.
Therefore, the accurate division of LF-HF information and the precise learning of HF features are the keys to the design of frequency-domain SR network architecture. FNNSR and IFNNSR have limited reconstruction effects because they do not consider the division of LF-HF information. ORDSR applies a fixed threshold to divide the image spectrum into LF and HF spectrum, and multiple convolutional layers to extract HF features, which also has limited feature extraction performance. Our network architecture yields the finest performance through an adaptive feature division mechanism to retain the LF spectrum, and a feature recovery mechanism to extract multilevel features at arbitrary scale factors.

\begin{table}[!tb]
\caption{PSNR$\uparrow$/SSEQ$\downarrow$ results of arbitrary upscale factors on Set5 dataset, * denotes that the modules are retrained with our training dataset. The best results are in bold.}
    \label{tab:result2}
    \centering
    \begin{tabular}{|c|c|c|c|c|}
        \hline
         \diagbox{Methods}{Scale} & $\times$1.2 & $\times$1.5 & $\times$2.0 \\\hline
         $\text{OverNet \cite{behjati2021overnet}}^*$ &43.40/32.03 &40.32/33.48 &36.22/35.92 \\\hline
         $\text{ArbSR \cite{wang2021learning}}^*$ &43.42/33.27 &40.37/34.92 &36.59/36.85  \\\hline
         FreeSR &\textbf{43.89/29.54} &\textbf{40.41/31.65} &\textbf{36.83/34.66} \\\hline\hline
         \diagbox{Methods}{Scale} & $\times$2.3  & $\times$2.6 & $\times$3.0 \\\hline
         $\text{OverNet \cite{behjati2021overnet}}^*$ &36.15/40.26 &36.06/42.97 &34.55/44.28 \\\hline
         $\text{ArbSR \cite{wang2021learning}}^*$ &36.44/41.95 &36.12/43.52 &34.59/44.89 \\\hline
         FreeSR &\textbf{36.59/37.94} &\textbf{36.17/41.56}&\textbf{34.68/43.22} \\\hline\hline
         \diagbox{Methods}{Scale} & $\times$3.4  & $\times$3.7 & $\times$4.0 \\\hline
         $\text{OverNet \cite{behjati2021overnet}}^*$ &33.45/46.19 &32.31/49.65 &32.26/52.47 \\\hline
         $\text{ArbSR \cite{wang2021learning}}^*$ &33.49/46.82 &32.68/50.96 &32.46/53.14 \\\hline
         FreeSR &\textbf{33.60/45.74}& \textbf{32.92/48.72} &\textbf{32.51/51.53} \\\hline
    \end{tabular}
\end{table}

The qualitative results of different frequency-domain SR methods under scale factor $\times$4 are shown in Fig. \ref{fig:visualf}. We select images ``comic'' and ``baboon'' in Set14 to demonstrate the reconstruction effect. It can be observed from Fig. \ref{fig:visualf} that our method tends to produce clear edges and detailed texture. For example, the pendants on the neck and the beard of the baboon recovered from our method are sharper and closer to the texture structure of HR images than other methods. The PSNR results of FreeSR are better than others due to the precise retention of LF information and recovery of HF features.

The amplification performance at integer scale factors shows the superiority of our method in frequency-domain SR networks. We further verify the performance of the proposed method at arbitrary-scale factors. The existing arbitrary-scale SR networks \cite{hu2019meta, wang2021learning, behjati2021overnet,yun2022single} are all implemented in the spatial or spatial-spectral domain. We choose the SOTA arbitrary magnification network OverNet \cite{behjati2021overnet} and ArbSR \cite{wang2021learning} for comparison. These networks are retrained on the same dataset for a fair comparison. The PSNR/SSEQ results of arbitrary-scale factors on the Urban100 dataset are shown in Table \ref{tab:result2}. It can be observed that our method achieves better PSNR results compared with OverNet and ArbSR, and the SSEQ results are improved significantly. The qualitative results are shown in Fig. \ref{fig:3.5}, we choose two reconstructed images of scale factor $\times$3.5 from the Urban100 dataset to present the SR performance. The subjective effect on the image ``woman'' shows that FreeSR can generate clear eye edges and eye details, while OverNet and ArbSR intend to produce blurry and artificial textures. Our method is designed with a precise feature division and multi-level feature adaption, therefore can eliminate artifacts and learn accurate and realistic details.

The quantitative and qualitative results demonstrate the priority of our method in frequency and spatial domain SR networks. In real-world scenarios, the efficiency of the algorithm is also a very important factor in evaluating the performance of the algorithm. Therefore, we further evaluate the running time and parameters of different arbitrary-scale SR networks. Table \ref{tab:time} shows the average running time of images from the Set5 dataset. All the networks are implemented on the PyTorch platform, the run-time testing work is carried out on the MacOS M2 chip with an 8GB CPU. It can be observed that FreeSR can reconstruct SR images faster than ArbSR and OverNet. For example, at scale factor $\times$2.0, the average time to reconstruct an image is 23.378s for the ArbSR algorithm, 2.6425s for the OverNet algorithm, and only 0.47s for our method. In addition, there are differences in the running time of the spatial-domain arbitrary-scale SR methods (ArbSR, OverNet) at different scale factors. This is because ArbSR and OverNet have different input LR image sizes at different scale factors. The larger the scale factor, the smaller the input image size is, and thus the shorter the running time. Our FreeSR enlarges the LR image to the same size as the HR image before feeding it into the network. And the size of the image fed into the network is the same at different scale factors, so the running time is almost the same. As for the network parameters, OverNet implements a lightweight structure and thus has fewer parameters than other networks, and FreeSR has much fewer parameters than ArbSR. In summary, our approach can provide better reconstruction results while maintaining algorithmic efficiency.

\begin{table}[t!]
\caption{Average running time comparison on Set5 Dataset.}
\label{tab:time}
\centering
\begin{tabular}{|c|c|c|c|c|}
  \hline
scale    & ArbSR   & OverNet  & FreeSR     \\   \hline
$\times$2.0        & 23.378s & 2.6425s  & 0.4700s   \\  \hline
$\times$3.0        & 10.768s   & 1.3193s & 0.4698s   \\  \hline
$\times$4.0        & 6.093s  & 0.6609s   & 0.4699s   \\   \hline\hline
parameters         &40.4M  & 0.92M   & 3.23M   \\   \hline
\end{tabular}
\end{table}



%

\subsection{Discussion}

Although FreeSR is an efficient arbitrary-scale SR model, it still has some promotion space for its applicability. First, the image spatial-domain feature contributes a lot to the SR network. In the future, we will involve spatial features in this work to achieve a spatial-spectral feature cooperative SR network for better SR performance. Second, FreeSR currently only supports image arbitrary SR within a scale factor range $[1.1, 4.0]$. It is worthing to improve FreeSR for a larger and more practical scale factor range~\cite{lee2022local}. Third, the scene of real-time video compression is more urgent for image SR technology \cite{Yang2022CVPR, afonso2019video}, and video codec depends on many DCT operations. Hence, it is possible to integrate our FreeSR into the video codec framework to achieve efficient video compression.


\section{CONCLUSION}
\label{sec:conclusion}
In this paper, we propose FreeSR, a novel frequency-domain scale-aware image super-resolution network. By quantitatively analyzing the frequency spectra of LR-HR image pairs under different scale factors in the frequency domain, we prove that the LR image suffers from different degradations of high frequency information both regarding to scale factors and image content. We devise a scale-aware feature division module based on a deep reinforcement learning network, to realize a precise LF and HF spectra division. Moreover, we design a scale-aware feature recovery module in the frequency domain to capture multi-level features of arbitrary-scale factors. Experimental results show that our method outperforms the SOTA arbitrary-scale SR networks.

\bibliographystyle{IEEEtran}
\bibliography{IEEEabrv,main}
\end{document}